\documentclass[10pt]{wlscirep}
\usepackage[utf8]{inputenc}
\usepackage[T1]{fontenc}
\usepackage{xspace, xcolor, graphicx, txfonts, comment}
\usepackage{amsmath}
\newcommand{\etal}{\textit{et al.\@}\xspace}
\newcommand{\exvivo}{\textit{ex vivo}\xspace}

\newcommand{\invitro}{\textit{in vitro}\xspace}

\newcommand{\invivo}{\textit{in vivo}\xspace}

\newcommand{\enface}{\textit{en face}\xspace}

\newcommand{\um}{\(\muup\)m\xspace}
\newcommand{\cdeg}{$^\circ$C\xspace}
\newcommand{\OCDSl}{OCDS$_l$\xspace}

\graphicspath{{./}{./figures/}{./Figs/}}

\title{Image-based investigation of the zebrafish developmental process using \invivo dynamic and multi-contrast optical coherence tomography}

\author[1,$\dag$]{Cunyou Bao}
\author[1,$\dag$]{Shadil Basheer}
\author[3]{Aiyi Sui}
\author[1,2]{Ibrahim Abd El-Sadek}
\author[1]{Rion Morishita}
\author[1]{Yu Guo}
\author[1]{Shuichi Makita}
\author[3]{Makoto Kobayashi}
\author[1*]{Yoshiaki Yasuno}

\affil[1]{Computational Optics Group, University of Tsukuba, Tsukuba, Ibaraki, Japan}
\affil[2]{Department of Physics, Faculty of Science, Damietta University, New Damietta City, Damietta, Egypt}
\affil[3]{Department of Molecular and Developmental Biology, Institute of Medicine, University of Tsukuba, Tsukuba, Ibaraki, Japan}
\affil[*]{yoshiaki.yasuno@cog-labs.org}
\affil[$\dag$]{These authors contributed equally to this work.}

\begin{abstract} 
	We demonstrate \invivo dynamic optical coherence tomography (DOCT) imaging of zebrafish development from 2 weeks to 12 months post-fertilization, integrated with polarization-sensitive OCT (PS-OCT), OCT angiography (OCTA), and histological validation.
	Two DOCT algorithms were utilized: logarithmic intensity variance and late OCT correlation decay speed, which characterize the occupancy of dynamic scatterers and their motion speeds, respectively.
	Our results show that skin stripes exhibit high DOCT signals and it varies among the pigment-cell types.
	Furthermore, the combination of DOCT and PS-OCT captures the maturation of these stripes.
	In addition, DOCT and OCTA successfully visualized the developmental progression of blood and lymphatic vessels, as well as spinal tissues.
\end{abstract}

\begin{document}
	\maketitle
	\section*{Introduction}
Understanding the complex, dynamic processes that govern tissue development and maturation is a fundamental goal in biological studies.
	Zebrafish (Danio rerio), a well-established model organism, plays an essential role in this field \cite{bilotta2001, feitsma2008, brittijn2009}.
It possesses several key advantages over conventional mammalian models, including high reproductive capacity, rapid development, high genetic similarity to humans, and external development of its embryos.  
	These attributes enable the longitudinal investigation of its entire life cycle and the visualization of dynamic cellular and tissue processes within the developmental process \cite{Boppart1996, Glenn2002, Maack2004, yaniv2006, Mo2021}.

	To visualize these processes in zebrafish, several imaging techniques are commonly used \cite{liu2016, cutrale2019}. 
	Fluorescence-based imaging is widely applied in zebrafish studies\cite{ko2011}.
	Point-scanning fluorescence microscopy provides high molecular specificity via fluorescent labels and achieves subcellular-to-cellular resolution, making it well-suited for investigating particular biological processes. 
	For example, point-scanning fluorescence microscopy has been used to image coronary vessel regeneration after injury in transparent adult fish and to map embryonic membrane lipid order during early organogenesis\cite{ElSadek2020BOE, bakis2023}.
	Light-sheet fluorescence microscopy (LSFM) is another powerful fluorescence-based method that rapidly acquires three-dimensional (3D) images while reducing photobleaching and phototoxic effects compared with point-scanning methods under typical conditions\cite{pampaloni2015}. 
	These features support long-term \invivo imaging of zebrafish developmental processes. 
	Specifically, beating embryonic heart development, optic cup morphogenesis during eye development, and larval neutrophil distribution and inflammatory changes have been revealed by LSFM \cite{taylor2019,icha2016,logan2018}.
	In addition to fluorescence-based approaches, imaging modalities based on alternative contrast mechanisms have also been applied in zebrafish studies.
	Optoacoustic microscopy (PAM) provides a unique contrast based on optical absorption\cite{yao2013}. 
	This enables deep-tissue, label-free imaging of endogenous molecules like hemoglobin and melanin, thereby mitigating scattering-limited penetration in optical microscopy.
	Using PAM, developmental melanophore migration, deep 3D anatomy in opaque juvenile fish, and the 3D distribution of green-fluorescent-protein- (GFP-) labeled tissues in later-stage fish have been visualized \cite{kneipp2015, vetschera2023, omar2017}.
	
	For \invivo studies of zebrafish development, an ideal imaging modality is expected to support volumetric imaging of dynamic processes with sufficient penetration depth across developmental stages while providing informative contrast without exogenous labeling.
	In practice, fluorescence microscopy and LSFM offer label-based molecular specificity but require exogenous labels and can be challenged by reduced optical penetration caused by scattering\cite{vetschera2023, watkins2012, adjili2015}.
PAM provides label-free contrast from endogenous optical-absorbing tissues, but typically involves trade-offs between spatial resolution and imaging depth\cite{yao2013}.
	These considerations motivate a complementary imaging method that can provide label-free contrast, high penetration depth, and compatibility with longitudinal \invivo imaging.
	
	Optical coherence tomography (OCT) is an imaging technique based on low-coherence interferometry that offers a favorable balance of micrometer-scale resolution, millimeter-level penetration, and high imaging speed, making it a promising candidate for \invivo studies in zebrafish across multiple developmental stages.
	Because PAM signal originates primarily from optical absorption, PAM complements the scattering-based structural information obtained from OCT.
	To provide sensitivity beyond anatomical structures obtained from OCT, some functional contrast extensions have been developed. 
	For example, polarization-sensitive OCT (PS-OCT) analyzes the polarization states of the backscattered light and offers tissue polarization properties as additional information\cite{deBoer2017BOE}. 
	OCT angiography (OCTA) is another vasculature contrast technique, which assesses blood flow by analyzing the temporal variation in the OCT signal from the same location \cite{makita2006OpEx, mariampillai2008, RKWang2010JSQE, RKWang2010OL, YLJia2012OpEx}. 
	Conventional OCT and its contrast extensions (e.g., OCTA, PS-OCT) have been used to study a wide range of biological processes in zebrafish, from the formation of anatomical structures and microvasculature to tissue regeneration and pathology \cite{Boppart1996, bailey2012, chen2016, toms2019, alam2022, Lichtenegger2022JBO, Lichtenegger2022BOE, Lichtenegger2022SciRep, Lichtenegger2023BioEng, kim2023}.

	More recently, dynamic OCT (DOCT) has emerged as a new contrast extension \cite{Azzollini2023BOE, ChaoRen2024CB, Heldt2025BOE, Josefsberg2025BOE}.
	It specifically visualizes tissue and cellular activities by analyzing the temporal signal fluctuation captured during repeated OCT scans.
	The analysis is typically performed using methods such as the variance or standard deviation of the time-sequential OCT signals\cite{apelian2016, ElSadek2020BOE, thouvenin2017, Pircher2025BOE}, power spectral density\cite{Oldenburg2015Optica, munter2020, KYChen2024BOE, Pircher2025BOE}, autocorrelation analysis \cite{JHLee2012OpEx, ElSadek2020BOE}, or the phase difference between the complex OCT signals at two adjacent time points\cite{schulz2025}. 
This capacity makes DOCT uniquely suited for investigating dynamic processes that drive development. 
	To date, the applications of DOCT have been mainly focused on \invitro and \exvivo samples. 
	
DOCT has been applied to several types of samples, which includes \invitro samples such as cell cultures \cite{SHPark2021BOE}, structured models \cite{Oldenburg2012BOE, YuGuo2025arXiv}, organoids \cite{Scholler2020LSA, Morishita2023BOE}, and spheroids \cite{ElSadek2020BOE, ElSadek2021BOE, ElSadek2023SR, ElSadek2024SciRep, Swanson2026BOE}.
	The \exvivo applications have also been widely investigated \cite{thouvenin2017cell, munter2020, Munter2021BOE, Mukherjee2021SR, Mukherjee2022BOE, Musial2022TVST, Mukherjee2023SR, Pircher2025BOE}.
	There are also a few examples of \invivo DOCT applications  \cite{YuGuo2025arXiv,Pircher2025BOE,TXia2023Optica}, but applications to \invivo zebrafish studies are limited because of difficulties arising from the involuntary motion of the zebrafish, which generates artifactual DOCT signals.

	In this study, we addressed these challenges by applying a DOCT imaging protocol optimized for \invivo zebrafish imaging, which includes an optimized anesthetic protocol, sample fixation method, and signal processing.
	By combining DOCT with other functional contrasts, we aim to provide a multi-contrast assessment of the developmental processes of  vessels, spine-related tissues, and skin stripes, specifically in the tail region.

	\section*{Methods}
	
	\subsection*{Samples and study design}
	The AB strain of zebrafish (\textit{Danio rerio}) was used as the animal model for this study.
	All fish were raised and maintained in a standard recirculating aquaculture system under controlled conditions (water temperature of 28.5 \cdeg and a 14/10-hour light/dark cycle).
	
	For the developmental study, a single, age-matched group of zebrafish was established from a synchronized fertilization.
	Different individuals were sampled from this group at four distinct time points: 2, 3, 4, and 5 weeks post-fertilization (wpf). 
	At each stage, five fish were randomly selected for imaging (N = 5 per time point), for a total of 20 fish analyzed in this study as the development group. 
	Additionally, a separate group of adult fish was used as a mature reference. 
	This adult group consisted of five adult zebrafish at 12 months post-fertilization (mpf, N = 5).
	
	For \invivo imaging, zebrafish were first anesthetized by immersion in a 0.016\% (w/v) tricaine solution until cessation of opercular movement.
	The same anesthesia protocol was used throughout the study.
	Each fish was then transferred to a Petri dish containing the same anesthesia solution and positioned for measurement. 
	Following the measurements, the fish were transferred to fresh water and monitored until they made a full recovery.
	
	After the \invivo imaging conducted with the OCT system, the fish were processed for histological workup. 
	The samples were fixed in a 4\% paraformaldehyde (PFA) solution and then embedded in paraffin to create paraffin blocks. 
	Finally, the blocks were sectioned and stained with Hematoxylin and eosin (H\&E) for histological observation. 
	The micrographs were acquired with an inverted microscope (Olympus, CKX53) using 4$\times$ and 10$\times$ magnification objectives.
	
	\subsection*{System and scanning protocol}
	
	A customized swept-source Jones-matrix OCT system with a 1.3-µm wavelength was used for the zebrafish measurements. 
	The system has a scanning speed of 50,000 A-lines/s, accompanied by a lateral resolution of 18.1 \um, defined as the $1/e^2$ beam diameter at the beam waist, and an axial resolution of 14~\um in tissue.
	The details of this system were introduced elsewhere \cite{li2017, miyazawa2019}.
	
	The scanning range was set to 6 mm $\times$ 3 mm for the development group and was increased to 6 mm $\times$ 6 mm for the adult group to accommodate the larger trunk area after growth.
	All zebrafish were scanned using two scanning protocols. 
	For DOCT imaging, a 32-frame repeated raster-scanning protocol was conducted \cite{ElSadek2021BOE}.
	The transverse field was split into 8 sub-fields along the slow scan direction, and each sub-field consisted of 16 scan locations. 
	The raster scanning was repeated 32 times for each sub-field to acquire a time sequence of 32 cross-sectional frames; each frame consisted of 512 A-lines. 
	The inter-frame interval was 204.8 ms, resulting in a total acquisition time of 52.4 s for a 128-location volume. 
	
	In addition to the protocol for DOCT imaging, a standard raster-scan protocol with 4-frame repetition was applied to achieve OCT angiography (OCTA) and degree of polarization uniformity (DOPU) imaging. 
	This four-frame repeating protocol has a frame interval of 12.8 ms. The resulting volumetric dataset consisted of 512 $\times$ 128 $\times$ 402 pixels, corresponding to the fast-scan, slow-scan, and depth directions, respectively.
	
	\subsection*{Contrast generation}
	\label{sec:methodContrast}
	Five different contrasts were generated from the acquired OCT data. 
	The conventional scattering intensity image was generated by averaging the absolute-squared intensities of four polarization channels.
	
	For DOCT analysis, two DOCT contrast images were generated from the temporal sequence of the averaged intensity.
	The first one is logarithmic intensity variance (LIV)\cite{ElSadek2020BOE}, which is calculated as the time variance of the dB-scaled OCT time-sequence signal at each pixel.
	Since LIV is the variance of dB-scaled intensity, its unit is dB$^2$. A value of 0~dB$^2$ indicates no signal fluctuation (no dynamics).
	The second, late OCT correlation decay speed (\OCDSl)\cite{ElSadek2020BOE}, was computed. 
	It was defined as the temporal decorrelation rate of the OCT signal, i.e., the slope of the autocorrelation decay curve over a specific time delay range [204.8~ms, 1228.8~ms], which is an established definition and delay range presented in Ref.~\cite{ElSadek2020BOE}.
	As demonstrated by Feng \etal \cite{feng2025}, LIV is sensitive to the occupancy of dynamic scatterers, while \OCDSl is primarily sensitive to the rate of the moving scatterers, exhibiting a peak response within a specific range of motion speeds (denoted as the sensitivity window of \OCDSl).

	For visualization, pseudo-color images were generated by combining the OCT intensity and DOCT contrast (i.e., LIV and \OCDSl) using the HSV color space. 
	The LIV or \OCDSl values were linearly mapped to the Hue channel, while the mean OCT intensity was mapped to the Value channel. 
	The saturation was kept constant to maintain color purity.
	
	OCTA and DOPU images were also generated. 
	OCTA was calculated based on the complex correlation of the sequentially acquired four B-scans at each location\cite{Makita2016BOE}.
	DOPU, a PS-OCT contrast metric, is sensitive to the spatial uniformity of the backscattered light's polarization state\cite{Gotzinger2008OpEx}.
	In our specific implementation, DOPU with noise-correction was computed, with a spatial kernel of 3 $\times$ 3 pixels along fast-scan and depth directions\cite{Makita2014OL}.
	
	\subsection*{Motion correction}
	The DOCT scanning protocol required an imaging time of over 6 seconds for the 32- frame sequence at each location. 
	This long duration makes the measurements highly susceptible to bulk motion artifacts during \invivo imaging.
	Such motions cause erroneously high LIV values.
	Therefore, prior to DOCT image generation, bulk motion artifacts within each 32-frame temporal sequence were corrected using a phase cross-correlation algorithm. 
	For each sequence, the 16th frame served as the reference, and frame displacement relative to this reference was calculated using sub-pixel phase cross-correlation with a precision of 1/10th pixel. 
	The detected motion was subsequently corrected by applying sub-pixel shifts to the dB-scaled intensity images.
	 This algorithm was implemented using the SciPy and skimage libraries, as detailed in our previous work\cite{Morishita2023BOE}.
	A pre- and post-motion correction comparison is provided in Supplementary Material, Fig.~S7.
	
	\subsection*{Slab projection generation}
	To clearly visualize tissue activities at defined depths relative to the sample surface, an automatic pipeline to generate slab projection images was used.
	
	The pipeline first involved extracting the surface topography from the 3D OCT intensity volume. This extraction began with segmentation, where each intensity cross-sectional image was binarized to separate the sample from the background, using either a manually set global threshold or the automated Otsu’s method. 
	Subsequently, the resulting binary images were refined using a series of morphological filtering operations.
	This custom sequence involved a sequence of closing and opening operations with rectangular structuring elements, designed to remove extraneous pixels resulting from binarization noise and to connect discontinuous surface segments. 
	A surface-height map was generated by identifying the axial position of the most superficial pixel for each A-line in the refined cross-sectional images. 
	Finally, a 2D median filter was applied to the resulting surface-height map to remove outliers and ensure spatial smoothness.

	Slab projection images of each contrast of OCT intensity, LIV, \OCDSl, and DOPU were generated by averaging the pixel values within a defined axial depth range relative to the surface along the \enface plane.
	DOPU was included for the skin projections as an indicator of pigment distribution and stripe boundaries, and OCTA for the spinal-region projections to visualize major blood vessels alongside the DOCT signals.
	For the skin, relatively shallow slabs were utilized, spanning a depth range of approximately 4--5 pixels (around 30--36~\um) depending on the developmental stage. 
	For the spinal cord region, thicker slabs were employed, with depth ranges of approximately 30--60 pixels (around 220--440~\um), adjusted according to the developmental stage.
	The pseudo-color projection of each contrast was then generated by mapping the slab projection of LIV, \OCDSl, or DOPU to the Hue channel and mapping the intensity slab projection to the Value channel.

	For quantitative analysis, regions of interest (ROIs) containing 250 pixels were selected within the corresponding anatomical structures of each fish.
	The mean and standard deviation of the LIV and \OCDSl values within each ROI were calculated and used for quantitative comparison among different developmental stages and tissue regions.
	
	\section*{Results}
In this section, representative images at five time points (from 2 weeks post-fertilization to 12 months post-fertilization) are presented. 
	Subsequently, slab projection images of skin and spinal regions are presented for further investigation of specific developmental processes.
	
	\subsection*{Overall structural features at different developmental stages}
	\subsubsection*{The 2-wpf stage}
\begin{figure}
		\centering
		\includegraphics[width=\linewidth]{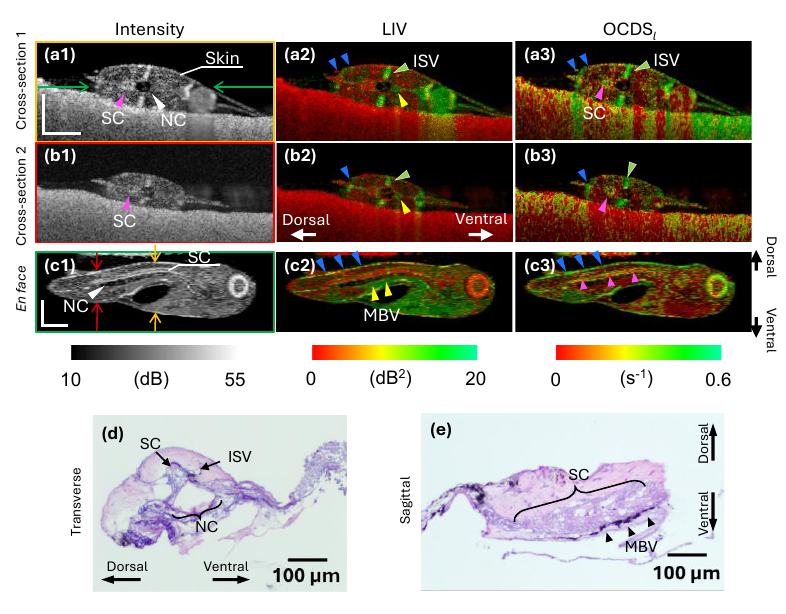}
		\caption{
			Cross-sectional and \textit{en face} images of a 2-wpf zebrafish.
			(a1)--(a3) OCT intensity, LIV, and \OCDSl images at a location indicated by the orange arrow pair in (c1).
			(b1)--(b3) OCT intensity, LIV, and \OCDSl images at a location indicated by the red arrow pair in (c1).
			(c1)--(c3) The corresponding \enface view of OCT Intensity, LIV, and \OCDSl. 
			The \enface slice is indicated by the green arrow pair in (a1).
			Colored outlines match the corresponding colored arrow pairs, indicating the same slice location. 
			This convention is used consistently in Figs.~1--4.
			(d)--(e) H\&E-stained histological sections of the trunk in transverse (d) and sagittal (e) views. 
			Scale bar represents 0.25 mm in (a1--c3).
			Abbreviation: SC, spinal cord; MBV, major blood vessels; NC, notochord; ISV, intersegmental vessels.
			Arrowhead colors indicate specific anatomical structures, consistent throughout the figures.
			Supplementary Figure S1 presents the identical images shown in the present figure, but without the overlaying labels.
		}
		\label{fig:slice_2wpf}
	\end{figure}
	Figure \ref{fig:slice_2wpf} presents multi-contrast cross-sectional and \enface images at 2 wpf compared with histology.
	Anatomical structures, such as the skin, spinal cord (indicated as SC), and notochord (indicated as NC), were identified in intensity images (a1--c1).
	The skin appeared as a thin hyper-scattering layer and exhibited high LIV and \OCDSl signals continuously along the dorsal side of the trunk [(a2--a3) and (b2--b3), blue arrowheads]. 
	The spinal cord was visualized as a rod-like structure dorsal to the notochord with high scattering intensity (a1, b1) and high \OCDSl signals [(a3, b3), pink arrowheads], whereas its LIV signal (a2, b2) remained low.
	Lateral to the notochord, vessel-like structures with both high LIV and \OCDSl values [(a2--a3) and (b2--b3), green arrowheads] were observed.
	These intramuscular vessel-like structures likely correspond to the intersegmental vessels (ISV).
	Notably, the dorsal aorta and posterior cardinal vein ventral to the notochord were exclusively highlighted by high LIV signals [(a2), (b2) and (c2), yellow arrowheads]; hereafter, these vessels are collectively referred to as the major blood vessels (MBV).
	In addition, tail artifacts beneath vessels were observed, as seen in Fig.~\ref{fig:slice_2wpf}(a2) and (a3).
	Notably, certain regions of the dish exhibit elevated LIV and \OCDSl signals.
	These artifacts do not represent true tissue dynamics but are caused by signal fluctuations originating from superior tissues.
	
	The corresponding H\&E-stained histology (d, e) also reveals the spinal cord, notochord, intersegmental vessels, and major blood vessels.
	Although the fish was small and fragile, leading to some structural distortion during histology preparation, the relative positions of the aforementioned tissues in the histological micrographs remain consistent with the OCT images.

	Supplementary Figure S1 presents the identical images as Fig.~ \ref{fig:slice_2wpf}, but without the overlaying labels.
	
	\subsubsection*{The 3-wpf stage}
\begin{figure}
		\centering
		\includegraphics[width=\linewidth]{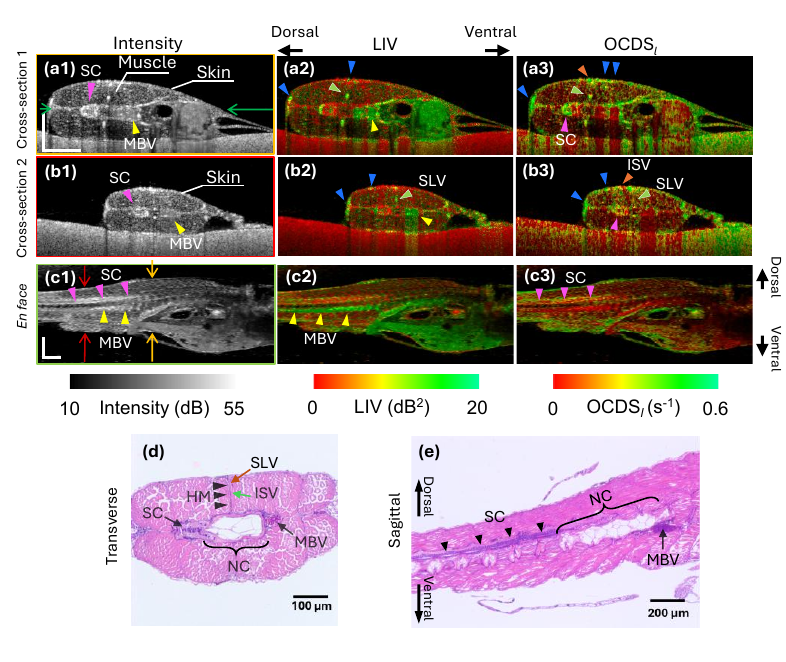}
		\caption{
			Cross-sectional and \enface images of a 3-wpf zebrafish.
			(a1)--(a3) OCT intensity, LIV and OCDS\textsubscript{\textit{l}} images at a location indicated by the orange arrow pair in (c1).
			(b1)--(b3) OCT intensity, LIV and OCDS\textsubscript{\textit{l}} images at a location indicated by the red arrow pair in (c1).
			(c1)--(c3) The corresponding \enface view of OCT Intensity, LIV and \OCDSl.
			The \enface slice is indicated by the green arrow pair in (a1).
			(d)--(e) H\&E-stained histological sections of the trunk in transverse (d) and sagittal (e) views.
			Scale bar represents 0.5 mm in (a1--c3).
			Abbreviation: SC, spinal cord; MBV, major blood vessels; NC, notochord; ISV, intersegmental vessels; SLV, superficial lymphatic vessels; HM, horizontal myoseptum.
			Supplementary Figure S2 presents the identical images shown in the present figure, but without the overlaying labels.
		}
		\label{fig:slice_3wpf}
	\end{figure}
	At 3 wpf, the zebrafish exhibited significant developmental progression (Fig.~ \ref{fig:slice_3wpf}). 
	Overall trunk volume, muscle, and skin layers visibly increased in thickness in both intensity and histological images (a1, b1, c1, d and e). 
	Beyond the increase in body thickness, developmental progressions are also evident in different tissues. 
	The major blood vessel, which is discernible in intensity images (c1), continued to show high LIV signals [(a2),(b2) and (c2), yellow arrowheads]. 
	The spinal cord maintained its high intensity and high \OCDSl signals (a1, a3, b1 and b3, pink arrowheads), with the latter appearing continuous in the \enface view (c3). 
	Furthermore, high LIV (a2, b2) and \OCDSl (a3, b3) signals emerged within the skin layer (blue arrowheads), a feature not observed at 2 wpf (Fig.~ \ref{fig:slice_2wpf}).
	
	In addition, a key developmental change was the segregation of the intramuscular vasculature, observed in DOCT images (a2, a3, b2 and b3) into two distinct vessels: the intersegmental vessel (indicated as ISV, green arrowheads) in the deep position and the superficial lymphatic vessel (SLV, orange arrowheads).
The intersegmental vessel (deeper vessel) exhibited both high LIV and high \OCDSl signals. 
	In contrast, the superficial lymphatic vessel showed high \OCDSl but moderate LIV signals, suggesting a smaller occupancy of dynamic scatterers compared to the intersegmental vessel. 
	The histological images (d, e) confirmed this segregation, showing the intersegmental vessel and superficial lymphatic vessel distributed along the horizontal myoseptum (HM) at shallow and deep layers, respectively.
	
	These distinctive DOCT signal characteristics might reflect the underlying physiological difference between the two vascular systems.
	The superficial lymphatic vessel is a low-pressure vessel responsible for transporting immune cells and interstitial fluid. 
	In contrast, the intersegmental vessels constitute a high-pressure system that carries blood to supply the skin and muscles\cite{cha2012,jung2017}.
	The functional DOCT images successfully captured this segregation during development. 
	A detailed discussion on the correlation between DOCT signals and the dynamics in superficial lymphatic vessels and intersegmental vessels is given later in \textit{Dynamics in the lymphatic system development} section.
	
	Supplementary Figure S2 presents the identical images with Fig.~ \ref{fig:slice_3wpf}, but without the overlaying labels.

	\subsubsection*{The 4- and 5-wpf stages}
\begin{figure}
		\centering
		\includegraphics[width=\linewidth,height=0.62\textheight,keepaspectratio]{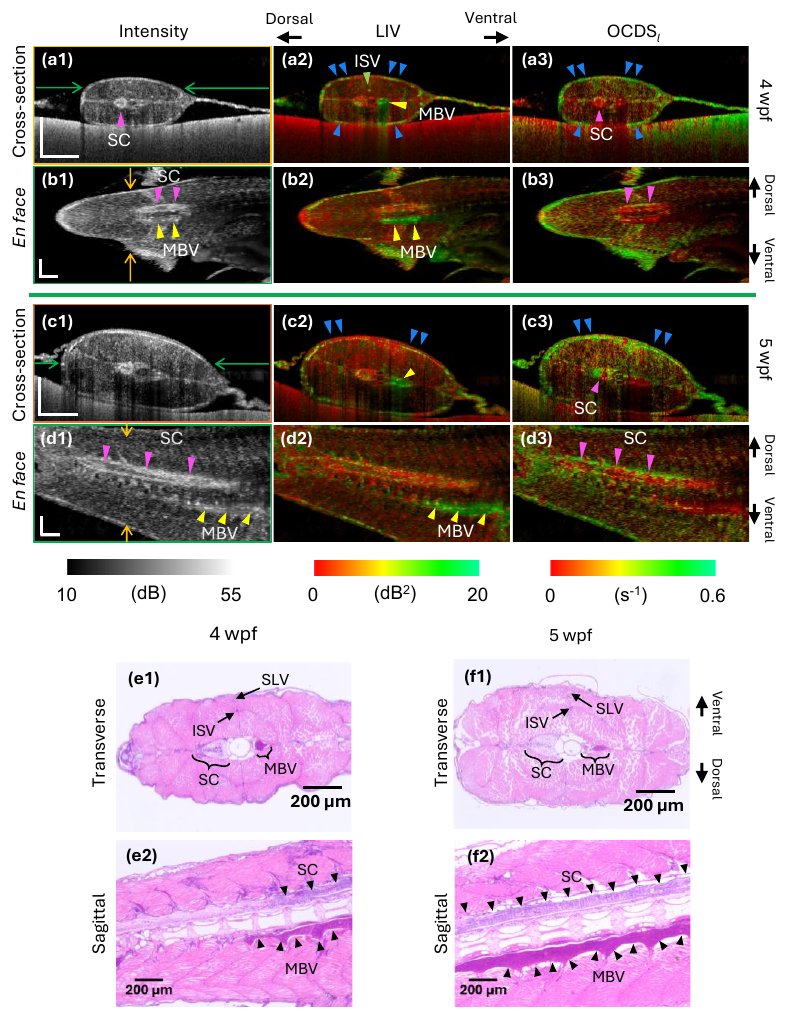}
		\caption{
			Cross-sectional and \enface images of 4-wpf and 5-wpf zebrafish.
			(a1)-(a3) OCT Intensity, LIV and \OCDSl images from a cross-section of the posterior trunk in 4-wpf zebrafish.
			(b1)-(b3) The corresponding \enface view of OCT Intensity, LIV and \OCDSl in the 4-wpf zebrafish.
			(c1)-(c3) OCT Intensity, LIV and OCDS\textsubscript{\textit{l}} images from a cross-section of the posterior trunk in 5-wpf zebrafish.
			(d1)-(d3) The corresponding \enface view of OCT Intensity, LIV and \OCDSl in the 5-wpf zebrafish.
			(e1)--(f2) H\&E-stained histological sections of the 4- and 5-wpf zebrafish's trunk in transverse and sagittal  views.
			Color-coded arrow pairs indicate the spatial relationship between the cross-sectional and \enface images.
			Scale bar represents 0.5 mm in (a1--d3).
			Abbreviation: SC, spinal cord; MBV, major blood vessels; NC, notochord; ISV, intersegmental vessels.
			Supplementary Figure S3 presents the identical images shown in the present figure, but without the overlaying labels.
		}
		\label{fig:slice_45wpf}
	\end{figure}
	Figure \ref{fig:slice_45wpf} presented the results for 4- and 5-wpf zebrafish. 
Between these two stages, the primary change was volumetric organ growth, with no significant structural alterations observed, and therefore the results were summarized together.
	
	The dynamic signals of internal organs were evident in DOCT images. 
	Major blood vessels (indicated as MBV) were well-defined in intensity images as hyper-scattering regions [(b1, d1), yellow arrowheads] and maintained distinct high LIV signals [(a2, b2, c2 and d2), yellow arrowheads]. 
The spinal cord developed a hollow morphology (b1, d1), with high \OCDSl signals becoming notably concentrated along its dorsal aspect (pink arrowheads); this feature was visible in both cross-sectional and \enface views (a3, b3, c3 and d3).
	In contrast to the prominent visualization at 3 wpf, the intramuscular vasculature (the intersegmental vessels and superficial lymphatic vessels) was no longer clearly resolved (a2--a3 and c2--c3), likely due to increased muscle thickness.
	The morphology and spatial distribution of these organs were consistent with the corresponding histological images (e1--f2).
	
	A distinctive feature observed at both stages is the signal pattern within the skin layers. 
	The skin resolved into a layered structure, which was moderately visible at 4 wpf but distinct at 5 wpf, and exhibited specific high LIV and \OCDSl signals bilaterally (blue arrowheads). 
	This distribution of functional signals within the skin layers may be associated with the ongoing formation of characteristic zebrafish stripe patterns, which is discussed later in the \textit{Dynamics in stripe development} section.
	
	Supplementary Figure S3 presents the identical images with Fig.~ \ref{fig:slice_45wpf}, but without the overlaying labels.
	
	\subsubsection*{The 12-mpf stage}
	\begin{figure}
		\centering
		\includegraphics[width=\linewidth]{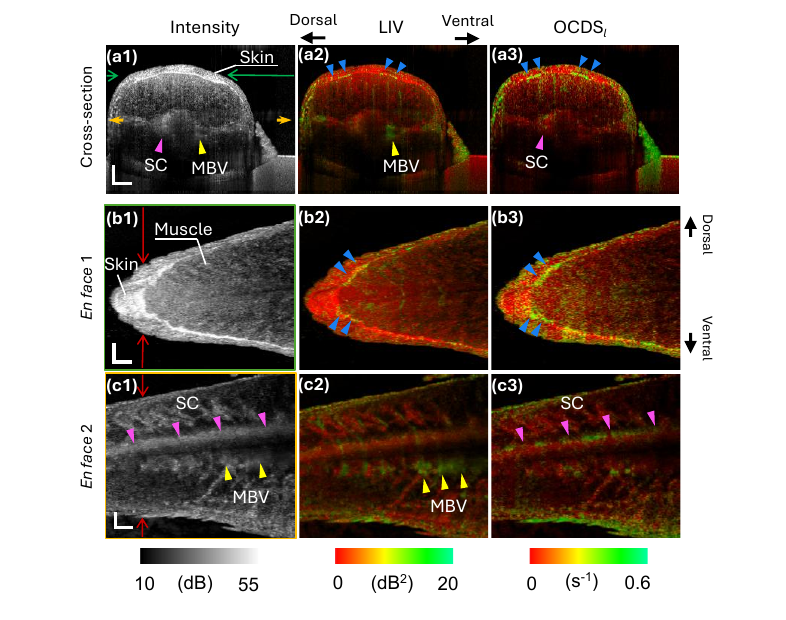}
		\caption{
			Cross-sectional and \enface images of the 12-mpf adult zebrafish trunk.
			(a1--a3) OCT Intensity, LIV and \OCDSl images from a cross-section of the posterior trunk in 12-mpf zebrafish.
			(b1--b3) and (c1--c3) show the \enface images from a superficial and a deep plane within the 12-mpf zebrafish, respectively.
			Color-coded arrow markers indicate the spatial relationship between the cross-sectional and \enface images.
			Scale bar represents 0.5 mm in (a1--c3).
			Abbreviation: SC, spinal cord; MBV, major blood vessels.
			Supplementary Figure S4 presents the identical images shown in the present figure, but without the overlaying labels.
		}
		\label{fig:slice_12mpf}
	\end{figure}
Figure \ref{fig:slice_12mpf} summarized the DOCT results of an adult zebrafish (12 mpf), which serves as a mature reference. 
	At this stage, the specific functional signatures observed in the late larval stages persisted, despite significant volumetric growth.
	The skin developed into a thick, stratified structure (a1) and maintained characteristic high LIV and \OCDSl regions [(a2--a3), blue arrowheads], which were clearly localized to the deepest skin layer and visible in the superficial \enface views (b2--b3).
	
	Internal organs also retained their distinct contrast features. 
	The enlarged spinal cord exhibited high \OCDSl signals (SC in a3, c3) concentrated on its dorsal aspect (pink arrowheads), consistent with the 4- and 5-wpf results. 
	Similarly, the major blood vessel was exclusively highlighted by high LIV signals (MBV, yellow arrowheads in c2) and low \OCDSl (c3), because the scatterers' speed inside exceeded the sensitivity range of \OCDSl. 
However, due to significant signal attenuation caused by increased tissue depth, structures beneath the spinal region could not be clearly resolved in the cross-sectional images (a1--a3).
	
	Supplementary Figure S4 presents the identical images with Fig.~ \ref{fig:slice_12mpf}, but without the overlaying labels.
	
	\subsection*{Slab projection analysis of key developmental events}
	The OCT and DOCT images at different developmental stages revealed several key developmental processes.
	Specifically, significant tissue activities related to stripe formation, vasculature maturation, and spinal cord development were observed. 
	Here we show a more detailed investigation of these events within the skin and spinal region using slab projection images.

	\subsubsection*{Projections of the skin region}
	\begin{figure}
		\centering
		\includegraphics[width=\linewidth]{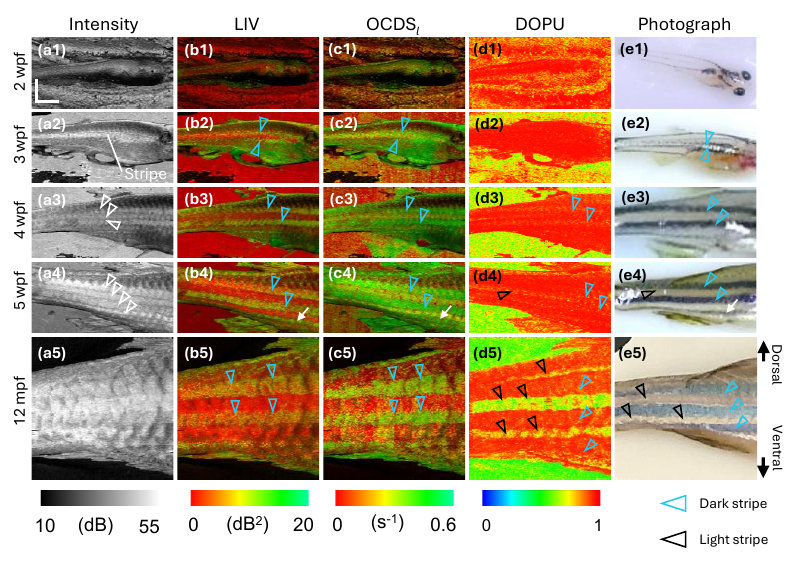}
		\caption{
			Multi-contrast slab projection images of zebrafish skin from 2-wpf to 12-mpf.
			Images were arranged in a grid, where each row represents a specific developmental stage (2-wpf to 12-mpf) and each column corresponds to a different imaging contrast. 
			And (e1--e5) are corresponding photographs for anatomical reference.
			Supplementary Figure S5 presents the identical images shown in the present figure, but without the overlaying labels.
		}
		\label{fig:slab_skin}
	\end{figure}
	Figure \ref{fig:slab_skin} summarized the time course of multi-contrast skin projections at all time points from 2 wpf to 12 mpf. 
	The characteristic dark- and light-stripe formation, which involves the precise arrangement of pigment cells and cellular interaction among different types of pigment cells, was visualized through the time course.
	
	During the early developmental stages (2--3 wpf), the zebrafish transitioned from transparent to semi-transparent (see (e1--e2)). 
	Sparse pigmentation initially appeared along the dorsal trunk and extended to the lateral side as diffuse clusters by 3 wpf (e1, e2). 
	This biological change was captured by the DOCT signals.  
	The 2-wpf larva showed sparse high-LIV signals only at the dorsal side, and distinct high LIV and \OCDSl signals began to organize into stripe-like patterns at 3 wpf [(b2, c2), blue arrowheads], correlating with emerging pigment clusters.
	
	By 4 to 5 wpf, the stripe patterns became more distinct. The dark stripes observed in photographs (e3, e4, blue arrowheads) were found to co-localize with the high signal regions in DOCT images (b3--c3, b4--c4, blue arrowheads), where dark-stripe regions consistently exhibited higher LIV and \OCDSl than inter-stripe regions at both stages (LIV and OCDS$_l$ values are summarized in Table \ref{tab:stripe_quant}).
	
	\begin{table}[ht]
		\centering
		\caption{DOCT signals (mean $\pm$ SD) in skin/stripe regions.}
		\label{tab:stripe_quant}
		\begin{tabular}{lcc}
			\hline
			& 4 wpf & 5 wpf \\
			\hline
			LIV, dark stripe (dB$^2$)         & 12.42 $\pm$ 3.03 & 10.77 $\pm$ 3.20 \\
			LIV, inter-stripe (dB$^2$)        & 3.49 $\pm$ 1.54  & 1.61 $\pm$ 0.99 \\
			OCDS$_l$, dark stripe (s$^{-1}$)  & 0.587 $\pm$ 0.062 & 0.542 $\pm$ 0.095 \\
			OCDS$_l$, inter-stripe (s$^{-1}$) & 0.355 $\pm$ 0.060 & 0.420 $\pm$ 0.081 \\
			\hline
		\end{tabular}
	\end{table}
	
	Notably, a new dark stripe emerged on the ventral side at 5 wpf (e4, white arrow), which was clearly resolved by both LIV and \OCDSl contrasts (b4--c4, white arrows).
	
	It is noteworthy that both the dark and light stripes exhibit increased scattering in the corresponding intensity images (a3, a4), while the inter-stripe regions appear as thin hypo-scattering lines.
The DOPU contrast (d3, d4) also began to resolve the stripe patterns.
	Namely, the dark stripes (blue arrowheads) exhibit low-DOPU at 4 wpf.
	At 5 wpf, the light stripe began to exhibit more distinct low DOPU lines (d4, black arrowhead) than the dark stripes.

	In the adult stage (12 mpf), the whole skin region exhibits high scattering (a5), where the dark and light stripe regions show slightly different scattering intensity, but the difference is not significant.
The DOPU images (d5) clearly visualized the stripe patterns, where the light (black arrowheads) and dark (blue arrowheads) stripes exhibit significantly and moderately low DOPU signals, respectively.

	These multi-contrast observations suggest distinct scattering and metabolic properties between dark and light stripes during development, which are further analyzed in \textit{Dynamics in stripe development} section.
	Supplementary Figure S5 presents the identical images with Fig.~ \ref{fig:slab_skin}, but without the overlaying labels.
	
	\subsubsection*{Projections of the spinal region}
	\begin{figure}
		\centering
		\includegraphics[width=\linewidth]{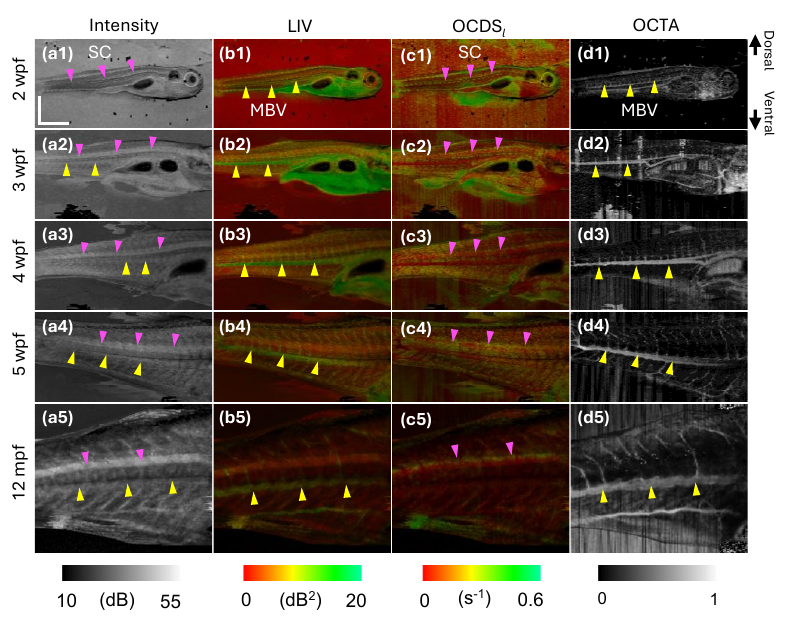}
		\caption{
			Multi-contrast slab projection images of the zebrafish spinal region from 2-wpf to 12-mpf.
			Images are arranged in a grid, where each row represents a specific developmental stage (2-wpf to 12-mpf) and each column corresponds to a different imaging contrast.
			Supplementary Figure S6 presents the identical images shown in the present figure, but without the overlaying labels.
		}
		\label{fig:slab_spine}
	\end{figure}
	Figure \ref{fig:slab_spine} summarizes the developmental changes in the spinal region from 2 wpf to adulthood. 
	The images clearly visualize the primary axial structures: the spinal cord (indicated as SC, pink arrowheads) and major blood vessels (MBV, yellow arrowheads) (a1--a5).
	
	Throughout the developmental time course, the vascular components exhibited consistent contrast features. 
	The major blood vessels, comprising the dorsal aorta and posterior cardinal vein, were consistently highlighted by high signals in both LIV and OCTA images [(b1--b5) and  (d1--d5), yellow arrowheads], revealing the formation of the complex vascular system during development, with LIV remaining high and \OCDSl remaining low throughout 3 to 5~wpf (Table \ref{tab:mbv_spine_quant}).
	
	In contrast, the spinal cord exhibited a distinct signal change during its development.
	At the early stage (2 wpf), high \OCDSl signals were distributed throughout the entire spinal cord structure [(c1), pink arrowhead]. 
	As development progresses, the high \OCDSl signal region began to localize to the dorsal aspect of the spinal cord (c2--c5), with mean spinal \OCDSl declining progressively from 3 to 5~wpf (summarized in Table \ref{tab:mbv_spine_quant}), reflecting the progressive restriction of high-\OCDSl signal to a shrinking dorsal region.

	\begin{table}[ht]
		\centering
		\caption{DOCT signals (mean $\pm$ SD) in major blood vessels (MBV) and spinal cord.}
		\label{tab:mbv_spine_quant}
		\begin{tabular}{lccc}
			\hline
			& 3 wpf & 4 wpf & 5 wpf \\
			\hline
			LIV, MBV (dB$^2$)         & 14.34 $\pm$ 2.39 & 12.82 $\pm$ 0.79 & 11.49 $\pm$ 1.01 \\
			LIV, spine (dB$^2$)       & 6.95 $\pm$ 1.53  & 5.77 $\pm$ 0.60  & 4.11 $\pm$ 0.59 \\
			OCDS$_l$, MBV (s$^{-1}$)  & 0.060 $\pm$ 0.024 & 0.041 $\pm$ 0.018 & 0.086 $\pm$ 0.030 \\
			OCDS$_l$, spine (s$^{-1}$)& 0.143 $\pm$ 0.042 & 0.110 $\pm$ 0.041 & 0.091 $\pm$ 0.031 \\
			\hline
		\end{tabular}
	\end{table}
	
	This observed shift likely reflects the underlying neuronal organization along the dorsal-ventral axis. 
	The spinal cord is known to contain different cell types located in specific domains along the dorsal-ventral axis \cite{lewis2003,cucun2024}. 
	The restriction of \OCDSl signals to the dorsal region in mature stages suggests that specific neuronal or glial activities in the dorsal domain maintain high intracellular dynamics, distinguishing them from the ventral regions during maturation.
	A more detailed discussion is given later in \textit{Dynamics in the spine} section.
	Supplementary Figure S6 presents the identical images with Fig.~ \ref{fig:slab_spine}, but without the overlaying labels.
	
	Table~\ref{tab:stage_summary} summarizes the structural and DOCT signal changes observed across all developmental stages.
	
	\begin{table}[ht]
		\centering
		\caption{Summary of structural features and DOCT signal characteristics of zebrafish skin, vasculature, and spinal cord across developmental stages.}
		\label{tab:stage_summary}
		\begin{tabular}{p{1.5cm}p{4cm}p{4.5cm}p{4.5cm}}
			\hline
			\textbf{Stage} & \textbf{Structural features} & \textbf{Skin and pigment DOCT signal} & \textbf{Vasculature and spinal cord DOCT signal} \\
			\hline
			2 wpf & Thin skin; spinal cord and major blood vessels visible adjacent to the notochord. & High LIV and OCDS$_l$ observed continuously along the dorsal skin. & Spinal cord: high OCDS$_l$, low LIV. ISV: high LIV and OCDS$_l$. MBV: high LIV. \\
			\hline
			3 wpf & Skin and muscle thickness increase; intramuscular vasculature begins to segregate into superficial and deep vessels. & High LIV and OCDS$_l$ signals emerge within the skin, absent at 2~wpf. & ISV: high LIV and OCDS$_l$. SLV: moderate LIV, high OCDS$_l$. Spinal cord: sustained high OCDS$_l$. MBV: sustained high LIV. \\
			\hline
			4--5 wpf & Spinal cord develops a hollow morphology; intramuscular vasculature no longer clearly resolved. & Skin resolves into a layered structure with high LIV and OCDS$_l$ signals bilaterally. & MBV: high LIV. Spinal cord: high OCDS$_l$, concentrated along the dorsal aspect. \\
			\hline
			12 mpf & Thick, stratified adult skin; deeper structures obscured by increased tissue depth. & High LIV and OCDS$_l$ localized to the deepest skin layer. & Spinal cord: high dorsal OCDS$_l$. MBV: high LIV, low OCDS$_l$. \\
			\hline
		\end{tabular}
	\end{table}
	
	\section*{Discussion}
	The period between 2 and 5 wpf is a key developmental stage for zebrafish, known as the larval-to-juvenile transition. 
	During this time, the fish's body changes dramatically to build its adult form. 
	On the outside, protective scales are formed, and the larval pigment pattern rearranges to form the distinct horizontal stripes running laterally along the fish's length\cite{singleman2014,patterson2019}. 
	Inside, the soft skeleton hardens into bone, making the fish stiff. 
	The vascular and lymphatic systems undergo a highly analogous process of maturation and remodeling. 
	Both vascular and lymphatic systems experience network refinement and stratification, transforming their initial, diffuse vessels into more hierarchical structures\cite{jung2017,cha2012,yaniv2006}.
	These changes are all essential for preparing the fish for adult life.
	
	In this study, we have successfully demonstrated a multi-contrast platform, including dynamic OCT, for the longitudinal, \invivo, label-free tracking of zebrafish development.
	Our work enabled the simultaneous visualization and differentiation of distinct physiological dynamics, such as the development of the vasculature, growth of the spinal cord, and stripe formation.
	In the following subsections, we explain our findings in relation to well-known key structural developmental events in zebrafish.
	
	Tricaine was used to immobilize the fish during imaging and may influence physiological flow; previous studies have shown that tricaine can affect cardiac function in zebrafish, particularly at higher concentrations, with effects depending on developmental stage and exposure conditions\cite{huriaux2010,wang2017}. A detailed investigation of the exact influence of tricaine on longitudinal developmental studies remains an important direction for future work. Notably, combining optical clearing with immunohistochemistry can provide functional contrast in deep tissues\cite{Affaticati2017zPACT}. Such an approach would complement dynamic OCT imaging, particularly for characterizing specific disease models.
	
\textbf{Dynamics in the lymphatic system development:}
	During the early developmental stage, the superficial lymphatic vessels and intersegmental vessels undergo rapid structural and functional differentiation while maintaining close spatial proximity along the horizontal myoseptum\cite{cha2012}.
	Anatomically, despite their close location, the internal fluid dynamics differ significantly: the intersegmental vessels carry blood at moderate velocities, whereas the superficial lymphatic vessels transport lymph fluid at a lower speed\cite{cha2012,jung2017}. 
	At the 2-wpf stage, DOCT images revealed a vessel-like structure exhibiting high LIV and \OCDSl signals in the muscle [Fig.~ \ref{fig:slice_2wpf}]. 
	By 3 wpf, this signal resolved into two vessel-like structures at two different image depths, corresponding to the locations of the superficial lymphatic vessel and intersegmental vessel identified in the anatomical images [Fig. \ref{fig:slice_3wpf}].
	
	The DOCT signals in these structures are consistent with the expected flow characteristics, as summarized in Table~ \ref{tab:vessels_signal}.
	The intersegmental vessel exhibited high LIV and \OCDSl values.
	The high LIV potentially reflects a high fraction of moving scatterers, and the high \OCDSl values suggest that the flow speed of the scatterers fell within the sensitivity window of \OCDSl (see the \textit{Contrast generation} section in Methods).
\begin{table}[ht]
		\caption{DOCT signal levels in major blood vessels (MBV), intersegmental vessels (ISV), and superficial lymphatic vessels (SLV) at 2 and 3 wpf.
		}
		\label{tab:vessels_signal}
\centering
		\begin{tabular}{lll}
			\hline
			Vessel type & LIV & \OCDSl\\
			\hline
			Major blood vessels (MBV) & High & Low \\
			Intersegmental vessels (ISV) & High & High \\
			Superficial lymphatic vessels (SLV) & Moderate & High \\
			\hline
		\end{tabular}
	\end{table}
	
	In contrast, the major blood vessels (dorsal aorta and posterior cardinal vein) did not show high \OCDSl signals [Figs.~ \ref{fig:slice_2wpf} and \ref{fig:slice_3wpf}], likely because their faster blood flow speeds exceeded the upper limit of the \OCDSl sensitivity window.
	The dynamic signals in the superficial lymphatic vessel (moderate LIV and high \OCDSl) may originate from two sources.
	First, although lymphatic flow is slow, the occupancy of the moving immune cells and other scatterers in the fluid can be high, and it results in high LIV. 
	In addition, the speed of the lymphatic flow may have fallen within the sensitivity window of \OCDSl.
Second, at early stages, the superficial lymphatic vessel underwent active assembly and remodeling during this developmental stage, which aligned with prior developmental studies on the lymphatic systems of zebrafish\cite{jung2017}.
	Such structural reorganization likely produces transient intracellular or subcellular motion, captured as moderate LIV and high \OCDSl signals.
	
	After 4 wpf, both the superficial lymphatic vessel and intersegmental vessel appeared less distinct in DOCT images.
	We attribute this reduction in visibility to two factors. 
	First, the developmental pace of lymphatic network formation slowed down following the larval-to-juvenile transition, and vessel remodeling began to stabilize\cite{jung2017}.
	Second, progressive thickening of the skin and musculature increased optical attenuation, thus reducing the sensitivity of DOCT images to dynamic processes within small, deep-tissue structures. 
	
	These findings highlight DOCT's ability to detect transient vascular and lymphatic dynamics associated with early developmental processes.
	By providing sensitivity to diverse motion sources, such as low-speed lymphatic flow and remodeling-related scatterer dynamics, DOCT offers complementary information that enables \invivo visualization of developmental processes.
	
\textbf{Dynamics in stripe development:}
	Zebrafish possess alternating dark and light horizontal stripes, resulting from the precise arrangement of three main classes of pigment cells: black melanophores, yellow xanthophores, and iridescent iridophores. 
	The establishment of this pigment pattern relies on the tight coordination of pigment cell differentiation, migration, proliferation, and cellular interaction\cite{patterson2019,nusslein2017}.
	
	During the formation of stripes, a distinct pattern of high LIV and \OCDSl signals consistently appeared within the presumptive and established dark stripe regions [Fig. \ref{fig:slab_skin}].
	These DOCT signals likely originated from the characteristic activities of melanophores.
	Melanophores expand dramatically during stripe formation, undergoing changes in cell size, dendritic remodeling, and pigment distribution\cite{patterson2019}.
	In contrast, the DOCT signals detected in adult zebrafish are unlikely to originate from developmental remodeling, as the pigment pattern has already been established.
	The high DOCT signals in adult dark stripes are more plausibly related to the high intracellular motility of melanophores.
	Melanosomes in the melanophores are known to have high motility of aggregation–dispersion for background-light adaptation \cite{logan2006}.
The aggregation of these organelles (i.e., melanosomes) in the cell center lightens the skin, while their dispersion darkens it.
	The high motility of the melanosomes may have caused the high LIV and \OCDSl signals in the dark stripe regions.

	In other contrasts, the increased scattering intensity observed in dark stripes may reflect the high concentration of melanin within melanophores. 
	The DOPU contrast exhibited an expanding pattern with low-DOPU during development, first appearing in the dark stripe regions and subsequently extending into the light stripes.
	The uniform low-DOPU in light stripes is consistent with the dense arrangement of iridophores, which are known to randomize the polarization state of backscattered light\cite{lichtenegger2022}. 
	The moderate reduction of DOPU in the dark stripe regions may be attributed to the depolarizing effect of melanin \cite{Baumann2012BOE}.

	The multi-contrast results indicate that the DOCT signals observed in stripe regions reflect dynamic biological processes associated with pigment cell behavior.
	While intensity and DOPU provide structural and polarization-based information related to pigment distribution and organization, DOCT is sensitive to the motility arising from cellular and intracellular activities during stripe formation.
	
\textbf{Dynamics in spine:}
	The spine in zebrafish is the central axial structure that provides primary support and protects the neural and vascular systems. 
	From 2 wpf onwards, the axial structures of the zebrafish trunk are well-organized. 
	In a typical cross-section, four primary structures are arranged sequentially from dorsal to ventral: the spinal cord, notochord, dorsal aorta, and posterior cardinal vein\cite{galbusera2019}.
	
	The major blood vessel, comprising the dorsal aorta and posterior cardinal vein, consistently exhibited high LIV signals throughout development.
	This is consistent with their function as vessels containing strong blood flow, indicating strong temporal fluctuations in the backscattered OCT intensity, and may have arisen from high dynamic scatterer occupancy and fast scatterer speed within these vessels.
	LIV, together with OCTA, reflected the hemodynamic activity within these vessels.
	Because the flow velocity exceeded the sensitivity window of \OCDSl, these vessels appeared as low-signal regions in \OCDSl images.

	In contrast, the spinal cord exhibited a distinct developmental pattern in \OCDSl images. 
	At early stages (2--3 wpf), high \OCDSl signals were distributed across the entire spinal cord.
	This uniform distribution may reflect globally active intracellular processes during early neural development.
	As development progressed, the high \OCDSl signal gradually became restricted to the dorsal aspect of the spinal cord.
	
	This spatial shift is consistent with the established dorsal-ventral patterning of the zebrafish spinal cord, where different neuronal populations occupy specific dorsal-ventral domains \cite{lewis2003,cucun2024}.
	The dorsal localization of \OCDSl signals at later stages, therefore, may indicate that intracellular activities remain more prominent in dorsal cell populations as development stabilizes.
	Although further investigation would be required to confirm these interpretations, the observed spatial shift of the \OCDSl pattern aligns well with established developmental patterns of the zebrafish spinal cord.
	
	\section*{Conclusion}
	We demonstrated \invivo DOCT and multi-contrast imaging to conduct a study of zebrafish development from 2 wpf to 5 wpf.
Our results demonstrate that DOCT effectively captures developmental processes at different stages.
	Specifically, these include the formation of dark stripes, the separation and remodeling of superficial lymphatic vessels and intersegmental vessels, and the growth of major blood vessels and the spinal cord.
	The combination of histological images and other functional OCT contrasts suggests that DOCT signals may indicate key events during development. 
	
	These results support the idea that the DOCT technique might work as a valuable tool to investigate and understand the developmental process in zebrafish.
	
	\section*{Funding}
	Japan Science and Technology Agency (JPMJCR2105); Japan Society for the Promotion of Science (21H01836, 22F22355, 22KF0058, 22K04962, 24KJ0510).
	
	\bibliography{ZebrafishDOCT}
	
	\section*{Author contributions}
	Bao, Kobayashi, Sui, and Yasuno designed the study.
	Kobayashi and Sui prepared the samples.
	Bao and Basheer organized the experiments and collected all the data.
	Bao, Basheer, El-Sadek, Morishita, Guo, and Makita analyzed and interpreted the data.
	Bao and Basheer wrote the first draft of the paper, and all authors revised the work and approved the final version of the manuscript.
	Yasuno supervised the work.
	
	\section*{Competing interests}
	Bao, Basheer, El-Sadek, Morishita, Makita, Yasuno: Nidek(F), Sky Technology(F), Panasonic(F), Nikon(F), Santec(F), Kao Corp.(F), Topcon(F). Sui, Kobayashi: None.
	
	\section*{Additional Information}
	Additional information can be found in the supplementary material.
	
	\section*{Data availability}
	Data underlying the results presented in this paper are not publicly available at this time but may be obtained from the authors upon reasonable request.
\end{document}